\documentclass[9pt]{article}
\usepackage{spconf, amsmath, graphicx}
\title{Efficient Target activity detection based on\\ recurrent neural networks}
\name{Daniel Gerber, Stefan Meier, and Walter Kellermann}
\address{Multimedia Communications and Signal Processing\\Friedrich-Alexander-University Erlangen-Nuremberg (FAU) \\
Cauerstr. 7, D-91058 Erlangen, Germany\\ 
  {\small \tt \{daniel.l.gerber, stefan.a.meier, walter.kellermann\}@fau.de}
  \vspace{.0cm}
}

\usepackage{url}
\usepackage{psfrag}
\usepackage{xcolor}
\usepackage{xspace}
\usepackage{float}
\usepackage[tight]{units}
\usepackage{caption}
\usepackage{subcaption}
\usepackage{setspace}
\usepackage{tabularx}
\usepackage{booktabs}
\usepackage{multirow}
\usepackage{bigdelim}
\usepackage{hhline}
\newcolumntype{Y}{>{\centering\arraybackslash}X}

\usepackage[acronym,nonumberlist]{glossaries}
\newacronym{bte}{BTE}{Behind-the-Ear}

\newacronym{pc}{PC}{Personal Computer}
\newacronym{ram}{RAM}{Random-Access Memory}
\newacronym{os}{OS}{Operating System}
\newacronym{cpu}{CPU}{Central Processing Unit}
\newacronym{gpu}{GPU}{Graphics Processing Unit}
\newacronym{cudnn}{cuDNN}{NVIDIA CUDA Deep Neural Network library}
\newacronym{cuda}{CUDA}{Compute Unified Device Architecture}

\newacronym{numpy}{NumPy}{NUMerical PYthon}
\newacronym{scipy}{SciPy}{SCIentific PYthon}

\newacronym{tad}{TAD}{Target Activity Detection}
\newacronym{sinr}{SINR}{Signal-to-Interference-plus-Noise Ratio}
\newacronym{doa}{DOA}{Direction Of Arrival}
\newacronym{vad}{VAD}{Voice Activity Detection}

\newacronym{gru}{GRU}{Gated Recurrent Unit}

\newacronym{lstm2}{LSTM}{Long Short-Term Memory}
\newglossaryentry{lstm}
{
  name={LSTM},
  description={Long Short-Term Memory},
  first={\glsentrydesc{lstm} (\glsentrytext{lstm})},
  plural={LSTMs},
  descriptionplural={Long Short-Term Memories},
  firstplural={\glsentrydescplural{lstm} (\glsentryplural{lstm})}
}

\newacronym{cec}{CEC}{Constant Error Carousel}

\newacronym{ctc}{CTC}{Connectionist Temporal Classification}

\newacronym{rnn}{RNN}{Recurrent Neural Network}
\newacronym{fnn}{FNN}{Feed-forward Neural Network}
\newacronym{mlp}{MLP}{Multilayer Perceptron}

\newacronym{hsrnn}{HSRNN}{Hierarchical Subsampling Recurrent Neural Network}

\newacronym{bp}{BP}{Backpropagation}
\newacronym{bptt}{BPTT}{Backpropagation Through Time}
\newacronym{sgd}{SGD}{Stochastic Gradient Descent}

\newacronym{roc}{ROC}{Receiver Operator Characteristics}
\newacronym{auc}{AUC}{Area Under Curve}
\newacronym{mcc}{MCC}{Matthew's Correlation Coefficient}
\newacronym{acc}{ACC}{Accuracy}
\newacronym{tpr}{TPR}{True Positive Rate}
\newacronym{fpr}{FPR}{False Positive Rate}
\newacronym{tp}{TP}{True Positives}
\newacronym{tn}{TN}{True Negatives}
\newacronym{fp}{FP}{False Positives}
\newacronym{fn}{FN}{False Negatives}

\newacronym{pdf}{PDF}{Probability Density Function}

\newacronym{mse}{MSE}{Mean Square Error}
\newacronym{ace}{ACE}{Averaged Cross-Entropy}
\newacronym{ssd}{SSD}{Sum Squared Differences}

\newacronym{smote}{SMOTE}{Synthetic Minority Over-sampling Technique}

\newacronym{kemar}{KEMAR}{Knowles Electronics Manikin for Acoustic Research}
\newacronym{matlab}{MATLAB}{MATrix LABoratory}

\newacronym{srp}{SRP}{Steered Response Power}
\newacronym{msc}{MSC}{Magnitude Squared Coherence}
\newacronym{rtt}{RTT}{Relative Training Time}

\makeglossaries

\usepackage{amsmath}
\usepackage{amsfonts}
\usepackage{amssymb}

\usepackage{bm}						   % matrix letters in math environment
\newcommand{\mat}[1]{\bm{#1}}          % ISO complying version

\newcommand{\figref}[1]{\hyperref[#1]{Fig.~\ref*{#1}}}
\newcommand{\tabref}[1]{\hyperref[#1]{Tab.~\ref*{#1}}}
\newcommand{\secref}[1]{\hyperref[#1]{Sec.~\ref*{#1}}}
\newcommand{\charef}[1]{\hyperref[#1]{Cha.~\ref*{#1}}}
\newcommand{\equref}[1]{\hyperref[#1]{\eqref{#1}}}

\usepackage{color} 
\definecolor{black}{gray}{0} 															
\usepackage[ colorlinks=true,
linkcolor=black,
citecolor=black,
breaklinks=true
]{hyperref}
\usepackage[all]{hypcap}	

\begin{document}
	%\ninept
	%
	\maketitle	
%
% SOF
%
\begin{abstract}
This paper addresses the problem of Target Activity Detection (TAD) for binaural listening devices.
TAD denotes the problem of robustly detecting the activity of a target speaker in a harsh acoustic environment, which comprises interfering speakers and noise (`cocktail party scenario').
In previous work, it has been shown that employing a Feed-forward Neural Network (FNN) for detecting the target speaker activity is a promising approach to combine the advantage of different TAD features (used as network inputs).
In this contribution, we exploit a larger context window for TAD and compare the performance of FNNs and Recurrent Neural Networks (RNNs) with an explicit focus on
small network topologies as desirable for embedded acoustic signal processing systems.
More specifically, the investigations include a comparison between three different types of RNNs, namely plain RNNs, Long Short-Term Memories, and Gated Recurrent Units.
The results indicate that all versions of RNNs outperform FNNs for the task of TAD.
%
%The work extends the classification scheme of Feed-forward Neural Networks (FNNs) to Recurrent Neural Networks (RNNs), in order to exploit the temporal evolution of the feature vectors.
%
\end{abstract}
%
% EOF
%

\begin{keywords}
	voice activity detection, target activity detection, recurrent neural networks, binaural listening devices
\end{keywords}

	%
%	\vspace{-.14cm}
	\section{Introduction}\label{sec:introduction}
%
% SOF
%
% \subsection{Target activity detection}\label{subsec:target_activity_detection}
%
% In the context of hearing aids, the problem of \gls{tad} is reflected in detecting the activity of a target speaker, in spite the presence of interferers and Noise, also referred to as `cocktail party scenario'. In binaural hearing aids, the left and right ear's information is exchanged via a wireless connection. This connection, in addition with several microphones per ear, allows for spatio-temporal filtering, e.g. beamforming. Since in an ordinary conversations humans tend to face each other, the speaker may be located in the line of sight of the listener, and therefore, the exploitation of the spatial information might be beneficial. With the acquired knowledge of the target activity, subsequent signal processing algorithms can benefit from this information via signal enhancement, adaptive filtering, etc.
%
Knowledge on the activity of a predefined target source is essential for many applications in speech signal processing.
%
%For many applications in speech signal processing, knowledge on the activity of a predefined target source is essential.
%
This knowledge can be exploited, e.g., in the context of automatic speech recognition, where the speech recognizer should only be active during target source activity \cite{ramirez2005effective}.
%
%Another application is the supervised estimation of relative transfer functions \cite{gannot2001signal}, e.g., binaural listening devices like hearing aids, where also reliable knowledge of time frames with a dominant target source is essential.
%Furthermore, the foregoing approach can also be generalized to the field of robot audition
%
%Another application is the supervised estimation of relative transfer functions. 
%They can be used, e.g., for signal enhancement in the context of binaural listening devices like hearing aids \cite{gannot2001signal} and in the field of robot audition \cite{7336933}.
%Since interfering sources degrade the estimation, a reliable detection of target activity becomes crucial.
%
%Since interfering sources degrade this estimation, a reliable detection of target activity becomes crucial.
%
Another application is the supervised estimation of Relative Transfer Functions (RTFs).
Since interfering sources have an impact on this estimation, a reliable detection of target activity becomes crucial.
The relative transfer functions can be used, e.g., for signal enhancement in the context of binaural listening devices like hearing aids \cite{gannot2001signal}.
Beyond this, RTFs can be extended to scatterer-related transfer functions for applications in the field of robot audition, e.g., \cite{barfuss2015hrtf}.
Relative impulse responses can in a subsequent step be used, e.g., for Linearly Constraint Minimum Variance (LCMV) beamforming \cite{hadad2016binaural}.
In this context, \gls{tad} is performed on embedded acoustic devices, where the need for computationally-efficient networks is most demanding. In order to achieve a comparable performance both on small network sizes and small amounts of training data, the selection of feature vectors is indispensable.
%
%
%
% EOF
%
%
% SOF
%
% \subsection{State of the art}\label{subsec:state_of_the_art}
%
Classical approaches for \gls{vad} are typically single-channel methods exploiting distinctive properties of speech signals like stationarity, harmonic structure and spectral envelopes in order to differentiate between speech and background noise \cite{graf2014improved,graf2015features}. These \gls{vad} methods, however, cannot be used to differentiate between a target speaker and interfering speech sources as the proposed \gls{tad} does. In this case, multichannel methods are beneficial, which can be based, e.g., on localization methods like the \gls{srp} method \cite{lee2009space,taghizadeh2011integrated}. In a similar way, the cross-correlation function between two microphones can be exploited for \gls{tad} by looking for peaks at the time lag corresponding to the (known) target source position \cite{denda2006robust,denda2007noise}, %\cite{koul2006using,denda2006robust,denda2007noise}
and the \gls{msc} allows for differentiating between a dominant coherent point source and incoherent background noise \cite{le1995study}. Moreover, it is possible to exploit conventional beamforming methods for \gls{tad} by estimating the \gls{sinr} based on one beamformer and one nullformer steered to the known target source \gls{doa}, yielding a target signal power estimate and a noise (and interference) power estimate, respectively \cite{herbordt2003acoustic,yu2010efficient}. %\cite{hoffman2001gsc,herbordt2003acoustic,yu2010efficient}.
Alternatively, monitoring the look direction of an adaptive nullsteering beamformer can provide information on target source activity \cite{srinivasan2008spatial}. A final group of \gls{tad} methods are probabilistic methods, which were also investigated in the recent past 
\cite{kim2008target,taseska2015minimum}.
%\cite{potamitis2004speech,kim2007voice,kim2008target,jarrett2014noise,taseska2015minimum}
%
% EOF
%
%
% \subsection{Goal}
%
% extend classical learning to sequence learning\\
% compare network types\\
% improve overall classification results
%
% \subsection{Outline}\label{subsec:outline}
%
Artificial neural networks were recently proposed for single-channel \gls{vad} \cite{graf2015features,wang2015universal,zhang2015boosting,moritz2016sprachaktiv}, %\cite{graf2015features,qi1993voiced-unvoiced,albu96application,wang2015universal,zhang2015boosting,moritz2016sprachaktiv}
but also for combining several features extracted by using information on the target source \gls{doa} \cite{meier2016interspeech}, outperforming the single-channel methods.
In this paper, the latter method will be extended to exploit a longer temporal context for \gls{tad} by using \glspl{rnn}.
\par
%
% \secref{sec:theory} covers the underlying theory of the extraction of the features, the process of sequence learning, and the different network topologies.
% The experiments section \secref{sec:experiments}, consists of a description of the implementation setup, as well as the obtained results.
% The investigated issues are briefly concluded in \secref{sec:conclusion}.
The remainder of this paper is organized as follows:
In \secref{sec:theory}, our \gls{tad} method is presented with focus on feature extraction, the process of sequence learning, and the different network topologies.
In \secref{sec:experiments}, the implementation setup is described and experimental evaluations are performed, leading to concluding remarks in \secref{sec:conclusion}.
	\vspace{-.2cm}
	\section{RNN-Based Target Activity Detection}\label{sec:theory}
	In this section, we provide details about the feature extraction (\secref{subsec:feature_extraction}), sequence learning (\secref{subsec:sequence_learning}), and the used network topologies (\secref{subsec:network_topologies}). For the clarity of presentation, we restrict our theoretical descriptions to network topologies with one hidden layer.
	\vspace{-.2cm}
%
%
% Definition of symbols for block index, feature vector and microphone signal
\newcommand\blockIndex{t}
\newcommand\featureSymb{f}
\newcommand\micSig[1]{\ensuremath{v_#1}\xspace}
\newcommand{\mejot}{\ensuremath{\mathrm{e}^{-\mathrm{j}\mu\Omega}}\xspace}
\newcommand\tphi{\ensuremath{\phi_\textrm{tar}}\xspace}
\newcommand\phidiff{\ensuremath{\phi_\textrm{diff}}\xspace}
\newcommand\rxx{\ensuremath r_{13}\xspace}
\newcommand{\fSNR}{\ensuremath{\featureSymb_\textrm{SNR}(\blockIndex)}\xspace}
\newcommand{\fdiff}{\ensuremath{\mat{\featureSymb}_\textrm{diff}(\blockIndex)}\xspace}
\newcommand{\fcorr}{\ensuremath{\featureSymb_\textrm{corr}(\blockIndex)}\xspace}
\newcommand{\fpos}{\ensuremath{\mat{\featureSymb}_\phi(\blockIndex)}\xspace}
\newcommand{\fvar}{\ensuremath{\mat{\featureSymb}_{\sigma^2}(\blockIndex)}\xspace}
\newcommand{\fcomb}{\ensuremath{\mat{x}_\blockIndex}\xspace}
\newcommand{\Dk}{\ensuremath{\Delta k}\xspace}
\newcommand{\Dkt}{\ensuremath{\Delta k_\textrm{tar}(\blockIndex)}\xspace}
\newcommand{\varmic}[1]{\ensuremath{\sigma_{\micSig#1}^2(\blockIndex)}\xspace}
\newcommand{\vars}{\ensuremath{\hat{\sigma}_\textrm{s}^2(\blockIndex)}\xspace}
\newcommand{\varn}{\ensuremath{\hat{\sigma}_\textrm{n}^2(\blockIndex)}\xspace}
\newcommand{\transp}[1]{\ensuremath{{#1}^\textrm{T}}\xspace}
\subsection{Feature extraction}\label{subsec:feature_extraction}
\begin{figure}
	\centering	
	\psfrag{x1}[bc][bc]{\small $\micSig1(k)$}
	\psfrag{x2}[bc][bc]{\small $\micSig2(k)$}
	\psfrag{x3}[bc][bc]{\small $\micSig3(k)$}
	\psfrag{x4}[bc][bc]{\small $\micSig4(k)$}
	\includegraphics[width=4cm]{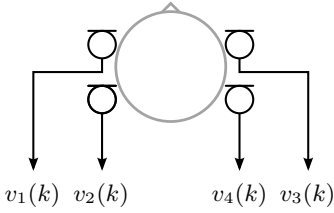}
	\caption{Sketch of the microphone positions.}
	\label{fig:micPositions}
\end{figure}
%
%
% data scenes, data basis\\
% feature extraction\\
% recursive averaging\\
% data sequencing
%
For each block $\blockIndex$, different features are calculated and stacked to a feature vector \fcomb (similar to \cite{meier2016interspeech, meier2016step}), which will be the input vector of the neural networks. We consider a microphone configuration as illustrated in Fig.~\ref{fig:micPositions}, where two pairs of microphones are placed on either side of a head (or more generally, at some distance on a scatterer), in order to both exploit the maximum aperture and allow for unilateral processing on the side of the target source.
\par
The first feature is an \gls{sinr} estimate obtained in a similar way as in \cite{herbordt2003acoustic,yu2010efficient,hoffman2001gsc} by steering a filter-and-sum beamformer and a nullformer (based on signals $\micSig1(k)$ and $\micSig3(k)$ in \figref{fig:micPositions}) in the known target source \gls{doa} \tphi in order to obtain estimates for target source power \vars and noise-plus-interference power \varn, respectively, yielding
\begin{equation}
	\fSNR = \frac{\vars}{\varn}.
\end{equation}
The second feature is based on $\rxx(\Dk,\blockIndex)$, which is the crosscorrelation function between $\micSig1(k)$ and $\micSig3(k)$ for block $\blockIndex$. Based on a known or estimated target source \gls{doa} $\tphi(\blockIndex)$, a time lag \Dkt can be determined, where the main peak of the target source is expected. This leads to the feature
\begin{equation}
	\fcorr = \frac{ r_{13}(\Dkt,\blockIndex)}{\max\limits_{\Dk \neq \Dkt} r_{13}(\Dk,\blockIndex)},
\end{equation}
which should exhibit a high value during target source activity. In \cite{srinivasan2008spatial}, a method for \gls{tad} was proposed, where an adaptive differential nullformer \cite{elko1995} minimizes its output power by adaptively steering a null into a direction $\phidiff(t)$. This beamformer uses the two microphones on the side of the head which is closer to the target source (i.e., either microphones 1 and 2, or microphones 3 and 4). If $\phidiff(t)$ is equal to $\tphi(t)$, this would give an indication for target activity and, hence,
\begin{equation}
	\fdiff = \transp{\left[\cos\left(\phidiff(\blockIndex)\right),\sin\left(\phidiff(\blockIndex)\right)\right]}
\end{equation}
forms a third component for the feature vector. Finally, the vectors
\begin{align}
	\fvar &= \transp{\left[\varmic1, \varmic3\right]}\\
	\fpos &= \transp{\left[\cos\left(\tphi(\blockIndex)\right), \sin\left(\tphi(\blockIndex)\right)\right]}
\end{align}
containing the powers of the microphone signals $\micSig1(k)$ and $\micSig3(k)$, and the sine and cosine of the known target source \gls{doa} $\tphi(t)$ are appended, which yields the feature vector
\begin{equation}\label{equ:feature_vector}
	\fcomb = \transp{\left[ \fSNR, \fcorr, \transp{\fdiff},  \transp{\fvar}, \transp{\fpos} \right]}.
\end{equation}
	\vspace{-.8cm}
%-----------------------------------------------------------------------------------------------------
% SOF
\subsection{Sequence learning}\label{subsec:sequence_learning}
%-----------------------------------------------------------------------------------------------------
%
Once the feature vector $\mat{x}_t$ is obtained, the classification task of \gls{tad} is performed, differentiating between activity and inactivity of the target speaker.
%
%\for{classical learning - fnn}
%
On the one hand, in a traditional approach, a \gls{fnn} serves as a memoryless classifier, where each feature vector $\mat{x}_t$ is mapped to the corresponding output vector $\mat{y}_t$, i.e., the classification result $\mat{y}_t$ is only based on the instantaneous observation $\mat{x}_t$ \cite{bishopnn}.
%
%\for{hidden state}
%
The hidden state vector $\mat{h}_t$ consists of the outputs of every neuron in the hidden layer and is related to the output vector $\mat{y}_t$ by the {\it softmax}-function \cite{graves_ssl}.
%, defined by
%\begin{align}
%y_{t,k} &= \frac{e^{a_{t,k}}}{\sum_{k=0}^{K-1}{e^{a_{t,k}}}}\\
%a_{t,k} &= \sum_{i}^{}{w_{ik}h_{t,i}+b_{k}},
%\end{align}
%where $k=0,\dots,K-1$, $K$ is the total number of classes, $b_{k}$ are scalar biases and $i$ is indexing the weights $w_{ik}$ and the components $h_{t,i}$ of the hidden state vector $\mat{h}_{t}$.
%\cite{graves_ssl}
%
%
\par
%
%\for{sequence learning - rnn}
%
%
On the other hand, the various versions of \glspl{rnn} exploit the temporal dependencies between subsequent feature vectors.
Each classification result is then dependent on previous hidden state vectors, as well as on the current input vector of the network.
While \glspl{fnn} can be trained on instantaneous feature vectors, \glspl{rnn} must be trained using sequences of feature vectors (`sequence learning').
%
%\for{figure description}
%
\begin{figure}
	\centering
	\scalebox{.9}{
	\psfrag{x0}[c][c]{\raisebox{-2mm}{\small$\mat{x}_0$}}
	\psfrag{x1}[c][c]{\raisebox{-2mm}{\small$\mat{x}_1$}}
	\psfrag{xn}[c][c]{\raisebox{-2mm}{\small$\mat{x}_{M-1}$}}
	\psfrag{h0}[c][c]{\raisebox{-2mm}{\small$\mat{h}_0$}}
	\psfrag{h1}[c][c]{\raisebox{-2mm}{\small$\mat{h}_1$}}
	\psfrag{hn}[c][c]{\raisebox{-2mm}{\small$\mat{h}_{M-1}$}}
	\psfrag{y0}[c][c]{\raisebox{-2mm}{\small$\mat{y}_0$}}
	\psfrag{y1}[c][c]{\raisebox{-2mm}{\small$\mat{y}_1$}}
	\psfrag{yn}[c][c]{\raisebox{-2mm}{\small$\mat{y}_{M-1}$}}
	\includegraphics[width=.28\textwidth]{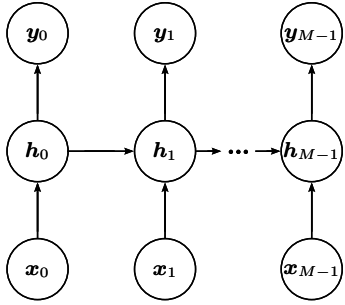}
}
	\caption{Schematic of sequence learning.}
	\vspace*{-.4cm}
	\label{fig:sequence_learning}
\end{figure}
\figref{fig:sequence_learning} shows the principle of sequence learning in an unrolled, or unfolded representation of the network over time.
The input sequence $\mat{x}_t$, starting with $\mat{x}_0$ up to $\mat{x}_{M-1}$, where $M$ is the sequence length, forming the context window of the network.
The hidden state vector $\mat{h}_t$ represents an \gls{rnn} layer of a given time step $t$,
$\mat{h}_{t+1}$ denotes the same layer on the next time step, and the horizontal arrow implies the recurrent connection between them.
The corresponding output sequence $\mat{y}_t$, i.e., $\mat{y}_0$ to $\mat{y}_{M-1}$, delivers the predictions of the network for the associated class labels.
%
%\for{sequence modes}
%
%For \glspl{rnn}, we consider two different types of mappings: The input sequence of feature vectors can either be mapped to a sequence of output vectors, or only to the last output vector $\mat{y}_{n-1}$.
%The latter case is representative for the sequence classification task of \gls{tad},
%where a series of feature vectors is considered for decision making.
%
In the sequence classification task of \gls{tad}, we consider the mapping of the input vectors to only the last output vector $\mat{y}_{M-1}$.
By assuming knowledge of the desired output, supervised learning of the neural networks is performed.
The \gls{bp} algorithm serves as a learning algorithm for \glspl{fnn} and \gls{bptt} is used for training \glspl{rnn} \cite{graves_ssl}.
%\gls{bptt} basically dissolves the recurrent connections by unfolding the network over time.
%The obtained network behaves deterministically and can be treated in a similar manner as \gls{bp} \cite{graves_ssl}.
%
%-----------------------------------------------------------------------------------------------------
% EOF
%-----------------------------------------------------------------------------------------------------
	\vspace{-.2cm}
%
% SOF
%
%-------------------------------------------------------------------------------------%
%
\subsection{Network types}\label{subsec:network_topologies}
For exploiting the temporal dependencies among the feature vectors according to \equref{equ:feature_vector} for the \gls{tad} task, we consider three types of neural networks with memory, namely plain \glspl{rnn} \cite{graves_ssl}, \glspl{lstm} \cite{hochreiter97}, and \glspl{gru} \cite{cho2014}, and compare them to the memoryless \glspl{fnn} \cite{bishopnn}.
The hidden state of the feed-forward layer is calculated as
\begin{align}\label{eq:fnn}		
\mat{h}_t = {f}(\mat{W}_{xh} \mat{x}_t + \mat{b}),
\end{align}
with the weight matrix $\mat{W}_{xh}$, the bias vector $\mat{b}$, and the nonlinear activation function ${f}$.
\equref{eq:fnn} is also referred to as gate, if it is employing a sigmoid function \cite{hochreiter97}.
In the following, for both \glspl{fnn} and \glspl{rnn} the function ${f}$ is chosen to be the hyperbolic tangent function $\tanh$ \cite{bishopnn}.
%
%
%-------------------------------------------------------------------------------------%
%
%\subsubsection{RNN}
%
As opposed to memoryless \glspl{fnn}, the group of \glspl{rnn} considers the temporal dependencies of subsequent feature vectors $\mat{x}_t$ by introducing recurrent connections to the previous time steps.
The hidden state vector of a plain \gls{rnn} is calculated as
\begin{align}
	\mat{h}_t={f}(\mat{W}_{xh} \mat{x_t} + \mat{W}_{hh} \mat{h}_{t-1} + \mat{b}), 
\end{align}
with the previous hidden state vector $\mat{h}_{t-1}$ weighted by the matrix $\mat{W}_{hh}$.
%
%
%-------------------------------------------------------------------------------------%
%
%\subsubsection{LSTMs and GRUs}
%
The inability of plain \glspl{rnn} to model long-time dependencies initially motivated the use of \glspl{lstm}, proposed by \cite{hochreiter97}, who introduced a memory unit, called cell state.
This cell state is accessed by gate units, limiting the effect of vanishing gradients \cite{hochreiter1998vanishing}.
\glspl{gru} were introduced to reduce the complexity of \glspl{lstm} while maintaining a similar expressive power, by dropping the memory unit and operating with gate units directly on the hidden state vector $\mat{h}_t$.
For a detailed description of the hidden state vector $\mat{h}_t$ of an \gls{lstm} see \cite{hochreiter97}, and for an \gls{gru} consider \cite{cho2014}.
	\vspace{-.05cm}
	\section{Experiments}\label{sec:experiments}
	We evaluated the proposed method in scenarios with up to five simultaneously active speech sources in reverberant environments for detecting time intervals, where the \gls{sinr} exceeds a threshold of \unit[10]{dB}, which is deemed relevant for practical applications \cite{haykin1996adaptive}.
	\vspace{-.2cm}
%-----------------------------------------------------------------------------------------------------
% SOF
\subsection{Implementation and scenarios}\label{subsec:implementation}
%-----------------------------------------------------------------------------------------------------
%
\begin{figure}
	\centering
	\includegraphics[width=.35\textwidth]{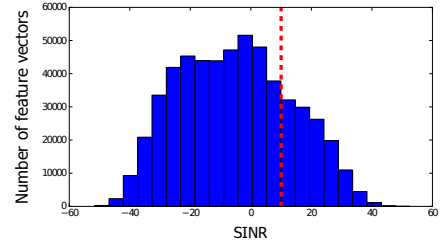}
	\caption{Histogram of input data SINRs of the training set in linear scale. Dashed line is indicating the threshold of \unit[10]{dB}.}
	\vspace{-.3cm}
	\label{fig:histogram}
\end{figure}
\begin{table*}[t]
	\centering

\begin{tabular}{c | c c c | c c c c c c}
	             &\multicolumn{3}{c|}{Performance} &\multicolumn{5}{c}{Complexity}  \\
	Network type & ACC & AUC & MCC                 & $N$ & $L$ & $P$ / $\overline{P}$ & RRT & RTT      \\
	\hline
	FNN (nos)	 & 0.801 & 0.906 & 0.539           & 32 & 6	& 5634 / 712	& 1.785	    & 1.188		  \\
	FNN (smo)	 & 0.870 & 0.950 & 0.662           & 32 & 2	& 1410 / 712	& 1		    & 1			  \\
	FNN (seq)	 & 0.889 & 0.950 & 0.700           & 32 & 6	& 10498 / 2308	& 1.0955	& 1.074		  \\
	RNN			 & 0.905 & 0.961 & 0.721           & 16 & 2	& 994 / 1545	& 14.021	& 9.221		  \\
	LSTM		 & 0.917 & 0.961 & 0.732           & 32 & 1	& 5474 / 7403	& 36.638	& 22.609	  \\
	GRU			 & 0.904 & 0.960 & 0.710           & 32 & 4	& 22850 / 5408	& 26.969	& 11.897	  \\
\end{tabular}

%-----------------------------------------------------------------------------------------------------
% EOT
%-----------------------------------------------------------------------------------------------------
%	\vspace{.2cm}
	\caption{
		Classification performance in terms of ACCuracy (ACC), Area Under the Curve (AUC) and Matthew's Correlation Coefficient (MCC), and complexity of the compared neural networks.
		$N$ denotes the number of neurons per layer, $L$ the number of layers, $P$ the total number of parameters,  $\overline{P}$ the average number of parameters over all 36 tested configurations per network type, RRT denotes the relative training time of the full training set, and RTT the relative testing time of the full test set.
	}
	\label{tab:result_complexity}
		\vspace*{-.4cm}
\end{table*}
For training the different network models and parameter sets, a software framework was implemented in Python, largely based on the library `Lasagne' \cite{lasagne}, which uses \gls{cuda} \cite{cuda}, performing the training using \gls{sgd} \cite{kingma2014adam}, an \gls{ace} cost function, and a {\it softmax} output layer \cite{graves_ssl}. For regularization, dropout \cite{dropout} is employed for \glspl{fnn} and synaptic noise \cite{synapticnoise} for \glspl{rnn}.
To keep the search space for the network parameters low, the number of hidden layers is varied from $L=1$ to $L=6$, and the number of neurons per layer is chosen between $N=1$ to $N=32$ in powers of two, which leads to a total of $36$ different configurations.
A batch of training data comprises $128$ sequences, with each sequence consisting of $20$ feature vectors.
The feature vector $\mat x_t$ is computed from the observed signals every millisecond according to \equref{equ:feature_vector}.
The dataset consists of recordings of a desired target speaker, up to $4$ simultaneously active interferers, and babble noise in the background.
The levels of targets and interferers are chosen to be equal and by varying the number of interferers different \glspl{sinr} are obtained from the recordings, with a background noise level at \unit[-10]{dB} relative to a speech source. The target and interferer positions are static, and varied in an angular range between \unit[-135]{$^{\circ}$} and \unit[+135]{$^{\circ}$}.
The speech sources were recorded at a distance of \unit[1]{m} in a living room-like environment at a sampling frequency of $f_s=\unit[16]{KHz}$. The scenarios are split into a set of $29$ acoustic scenes of length \unit[20]{s} (resulting in a total of $551,000$ labeled feature vectors) for training and validation, and $9$ acoustic scenes of length \unit[10]{s} (resulting in a total of $89,919$ labeled feature vectors) for testing purposes.
The ground truth for the target activity was defined by calculating the instantaneous \gls{sinr} (with knowledge of the individual target source and interferer components) and applying a threshold of \unit[10]{dB}, denoted as `\unit[10]{dB}-dataset'.
This leads to binary output values, which are used for supervised learning. \figref{fig:histogram} shows the \gls{sinr} distribution of the training data. The inequality in class labels is afterwards balanced by upsampling the minor class until equality in the number of class labels is reached.
The threshold of \unit[10]{dB} is chosen as a typical value for real-world requirements, as, e.g., Least Mean Squares (LMS)-type algorithms need a sufficiently high \gls{sinr} for convergence \cite{haykin1996adaptive}.
%
%-----------------------------------------------------------------------------------------------------
% EOF
%-----------------------------------------------------------------------------------------------------
	\vspace{-.4cm}
%-----------------------------------------------------------------------------------------------------
% SOF
\subsection{Results}\label{subsec:results}
%
% Measures
%-----------------------------------------------------------------------------------------------------
%
Six different network types are compared by their performance as well as their complexity in \tabref{tab:result_complexity}.
For each network type, the configuration (defined by the number of neurons per layer $N$ and the number of layers $L$) has been chosen in terms of \gls{mcc} \cite{fang2013classifying_performance_measures} on the validation set of the \unit[10]{dB}-dataset.
While \gls{mcc} and \gls{auc} \cite{fang2013classifying_performance_measures} of a receiver operator characteristics deliver a viable measure when applied to unbalanced data, the \gls{acc} measure \cite{fang2013classifying_performance_measures}, although commonly used, produces results of limited value.
The ratio $P/\overline{P}$ indicates the relative complexity of the chosen network compared to all other considered configurations.
%
%-----------------------------------------------------------------------------------------------------
% Complexity
%-----------------------------------------------------------------------------------------------------
%
`FNN (nos)' and `FNN (smo)' are examples for instantaneous learning, where a single feature vector is classified by an \gls{fnn}, exhibiting the lowest number of parameters on average.
`FNN (smo)' uses a recursively averaged feature vector $\overline{\mat{x}}_t$, given by $\overline{\mat{x}}_t = (1-a)\mat{x}_{t} + a \overline{\mat{x}}_{t-1},$
with $a=0.7$, whereas `FNN (nos)' uses the original feature vector ($a=0$).
`FNN (seq)', `\gls{rnn}', `\gls{lstm}' and `\gls{gru}' are examples of sequence learning approaches.
`FNN (seq)' employs an \gls{fnn} on a concatenation of a sequence of feature vectors $\mat{x}_{0}$ to $\mat{x}_{M-1}$, which are introducing roughly $M$ times more weights in the first layer than `FNN (nos)', where $M$ is the sequence length chosen to $M=20$.
In addition, the last three setups represent the group of \glspl{rnn}.
By additional recurrent connections, the plain \gls{rnn}, denoted as `\gls{rnn}', is only about twice as complex as an `FNN (nos)' on average, although performing sequence learning.
By introducing gate units to control the information flow inside a neuron, `\gls{lstm}' and `\gls{gru}' are the most demanding setups regarding the number of parameters.
%
%-----------------------------------------------------------------------------------------------------
% Results
%-----------------------------------------------------------------------------------------------------
%
\tabref{tab:result_complexity} indicates that by recursively averaging, a significant performance gain of `\gls{fnn} (smo)' over `\gls{fnn} (nos)' is observable, confirming the benefit of incorporating averaged feature vectors into the classification.
While `\gls{fnn} (seq)' outperforms the other \glspl{fnn} due to its larger number of inputs and, accordingly the larger number of parameters, all \glspl{rnn} outperform all considered feed-forward networks.
Especially, the plain \gls{rnn} requires roughly $10$ times less parameters compared to `\gls{fnn} (seq)', but delivers a better performance at the expense of an increased testing time.
The recurrent nature of the group of \glspl{rnn} indicate that they have learned the temporal evolution of the feature vectors through their feed-back connections.
\glspl{lstm} and \glspl{gru} are not able to benefit from their long-term memory, which may be due to the nonstationarity of speech signals.
While the three recurrent network types perform similarly well for \gls{tad}, the plain \gls{rnn} shows the lowest number of parameters, which renders it the model of choice, especially for embedded applications demanding for low complexity.
%
%-----------------------------------------------------------------------------------------------------
% EOF
%-----------------------------------------------------------------------------------------------------
	%
%	\vspace{-4pt}
%	\par
	%
	\vspace{-.1cm}
	\section{Conclusion}\label{sec:conclusion}
In this paper, a set of TAD features is used at the input of a neural network detecting the activity of a  desired speaker. As main innovation with respect to previous work, we propose to employ recursive layers in the neural network performing efficient TAD for embedded acoustic devices. In the experimental part using a multitude of challenging acoustic scenarios and comparing six different network types, we illustrate that RNNs outperform FNNs, pointing at the plain \gls{rnn} as the structure of choice, due to the lowest number of trainable parameters involved. Future work will include additional features to further improve the characterization of the scenarios.
%
%-----------------------------------------------------------------------------------------------------
% EOF
%-----------------------------------------------------------------------------------------------------
	\cleardoublepage
	%
% 	\vfill\pagebreak
% 	\clearpage
	{
	\small
	\begin{spacing}{0.91}
	\bibliographystyle{ieeetr}
	\bibliography{strings,refs}

\begin{thebibliography}{10}

\bibitem{ramirez2005effective}
J.~Ram{\'\i}rez, J.~C. Segura, C.~Ben{\'\i}tez, A.~De~la Torre, and A.~Rubio,
  ``An effective subband {OSF}-based {VAD} with noise reduction for robust
  speech recognition,'' {\em IEEE Trans. Speech Audio Process.}, vol.~13,
  no.~6, pp.~1119--1129, 2005.

\bibitem{gannot2001signal}
S.~Gannot, D.~Burshtein, and E.~Weinstein, ``Signal enhancement using
  beamforming and nonstationarity with applications to speech,'' {\em IEEE
  Trans. Signal Process.}, vol.~49, no.~8, pp.~1614--1626, 2001.

\bibitem{barfuss2015hrtf}
H.~Barfuss, C.~Huemmer, G.~Lamani, A.~Schwarz, and Kellermann, ``{HRTF}-based
  robust least-squares frequency-invariant beamforming,'' {\em Proc. IEEE
  WASPAA}, pp.~1--5, October 2015.

\bibitem{hadad2016binaural}
E.~Hadad, S.~Doclo, and S.~Gannot, ``The binaural {LCMV} beamformer and its
  performance analysis,'' {\em IEEE/ACM Trans. Audio, Speech, Language
  Process.}, vol.~24, no.~3, pp.~543--558, 2016.

\bibitem{graf2014improved}
S.~Graf, T.~Herbig, M.~Buck, and G.~Schmidt, ``Improved performance measures
  for voice activity detection,'' in {\em Proc. ITG Conf. Speech
  Communication}, pp.~1--4, VDE, 2014.

\bibitem{graf2015features}
S.~Graf, T.~Herbig, M.~Buck, and G.~Schmidt, ``Features for voice activity
  detection: a comparative analysis,'' {\em EURASIP J. Advances Signal
  Process.}, vol.~2015, no.~1, pp.~1--15, 2015.

\bibitem{lee2009space}
H.~Lee and D.~Yook, ``Space-time voice activity detection,'' {\em IEEE Trans.
  Consumer Electronics}, vol.~55, no.~3, pp.~1471--1476, 2009.

\bibitem{taghizadeh2011integrated}
M.~J. Taghizadeh, P.~N. Garner, H.~Bourlard, H.~R. Abutalebi, and A.~Asaei,
  ``An integrated framework for multi-channel multi-source localization and
  voice activity detection,'' in {\em Proc. IEEE Joint Workshop HSCMA},
  pp.~92--97, 2011.

\bibitem{denda2006robust}
Y.~Denda, T.~Nishiura, and Y.~Yamashita, ``Robust talker direction estimation
  based on weighted {CSP} analysis and maximum likelihood estimation,'' {\em
  IEICE Trans. Information and Systems}, vol.~89, no.~3, pp.~1050--1057, 2006.

\bibitem{denda2007noise}
Y.~Denda, T.~Tanaka, M.~Nakayama, T.~Nishiura, and Y.~Yamashita, ``Noise-robust
  hands-free voice activity detection with adaptive zero crossing detection
  using talker direction estimation,'' in {\em Proc. Annual Conf. Interspeech},
  pp.~222--225, 2007.

\bibitem{le1995study}
R.~Le~Bouquin-Jeann{\`e}s and G.~Faucon, ``Study of a voice activity detector
  and its influence on a noise reduction system,'' {\em Speech Communication},
  vol.~16, no.~3, pp.~245--254, 1995.

\bibitem{herbordt2003acoustic}
W.~Herbordt, H.~Buchner, and W.~Kellermann, ``An acoustic human-machine
  front-end for multimedia applications,'' {\em EURASIP J. Applied Signal
  Process.}, pp.~21--31, 2003.

\bibitem{yu2010efficient}
T.~Yu and J.~H. Hansen, ``An efficient microphone array based voice activity
  detector for driver's speech in noise and music rich in-vehicle
  environments,'' in {\em Proc. IEEE ICASSP}, pp.~2834--2837, IEEE, 2010.

\bibitem{srinivasan2008spatial}
S.~Srinivasan and K.~Janse, ``Spatial audio activity detection for hearing
  aids,'' in {\em Proc. IEEE ICASSP}, pp.~4021--4024, IEEE, 2008.

\bibitem{kim2008target}
H.-D. Kim, J.~Kim, K.~Komatani, T.~Ogata, and H.~G. Okuno, ``Target speech
  detection and separation for humanoid robots in sparse dialogue with noisy
  home environments,'' in {\em Proc. IEEE/RSJ Int. Conf. IROS}, pp.~1705--1711,
  IEEE, 2008.

\bibitem{taseska2015minimum}
M.~Taseska and E.~A. Habets, ``Minimum {B}ayes risk signal detection for speech
  enhancement based on a narrowband {DOA} model,'' in {\em Proc. IEEE ICASSP},
  pp.~539--543, IEEE, 2015.

\bibitem{wang2015universal}
Q.~Wang, J.~Du, X.~Bao, Z.-R. Wang, L.-R. Dai, and C.-H. Lee, ``A universal
  {VAD} based on jointly trained deep neural networks,'' in {\em Proc. Annual
  Conf. Interspeech}, pp.~2282--2286, 2015.

\bibitem{zhang2015boosting}
X.-L. Zhang and D.~Wang, ``Boosting contextual information for deep neural
  network based voice activity detection,'' {\em IEEE/ACM Trans. Audio, Speech,
  Language Process.}, vol.~24, no.~2, pp.~252--264, 2016.

\bibitem{moritz2016sprachaktiv}
N.~Moritz, J.~Drefs, H.~Baumgartner, and J.~Rennies,
  ``{S}prachaktivit{\"a}tserkennung basierend auf {D}eep {N}eural {N}etworks
  f{\"u}r {A}nwendungen in {F}ilm und {F}ernsehen,'' pp.~960--963, DAGA, March
  2016.

\bibitem{meier2016interspeech}
S.~Meier and W.~Kellermann, ``Artificial neural network-based feature
  combination for spatial voice activity detection,'' in {\em Proc. Annual
  Conf. Interspeech}, September 2016.

\bibitem{meier2016step}
S.~Meier and W.~Kellermann, ``Relative impulse response estimation during
  doubletalk with an artificial neural network-based step size control,'' in
  {\em Proc. IWAENC}, September 2016.

\bibitem{hoffman2001gsc}
M.~W. Hoffman, L.~Zhao, and D.~Khataniar, ``{GSC}-based spatial voice activity
  detection for enhanced speech coding in the presence of competing speech,''
  {\em IEEE Trans. Speech Audio Process.}, vol.~9, no.~2, pp.~175--179, 2001.

\bibitem{elko1995}
G.~Elko and A.-T.~N. Pong, ``A simple adaptive first-order differential
  microphone,'' in {\em Proc. IEEE WASPAA}, pp.~169--172, 1995.

\bibitem{bishopnn}
C.~M. Bishop, {\em Neural Networks For Pattern Recognition}.
\newblock Clarendon Press, 1996.

\bibitem{graves_ssl}
A.~Graves, {\em Supervised sequence labelling}.
\newblock Springer, 2012.

\bibitem{hochreiter97}
S.~Hochreiter and J.~Schmidhuber, ``Long short-term memory,'' {\em Neural
  computation}, vol.~9, no.~8, pp.~1735--1780, 1997.

\bibitem{cho2014}
K.~Cho, B.~Van~Merri{\"e}nboer, C.~Gulcehre, D.~Bahdanau, F.~Bougares,
  H.~Schwenk, and Y.~Bengio, ``Learning phrase representations using {RNN}
  encoder-decoder for statistical machine translation,'' {\em arXiv preprint
  arXiv:1406.1078}, 2014.

\bibitem{hochreiter1998vanishing}
S.~Hochreiter, ``The vanishing gradient problem during learning recurrent
  neural nets and problem solutions,'' {\em International Journal of
  Uncertainty, Fuzziness and Knowledge-Based Systems}, vol.~6, no.~02,
  pp.~107--116, 1998.

\bibitem{haykin1996adaptive}
S.~Haykin {\em et~al.}, ``Adaptive filtering theory,'' {\em Englewood Cliffs,
  NJ: Prentice-Hall}, 1996.

\bibitem{lasagne}
E.~Battenberg, S.~Dieleman, D.~Nouri, E.~Olson, A.~van~den Oord, C.~Raffel,
  J.~Schl\"uter, and S.~Sonderby, ``Lasagne: First release..''
  \url{http://dx.doi.org/10.5281/zenodo.27878}, Aug. 2015.
\newblock Retrieved on 2016-04-18.

\bibitem{cuda}
J.~Nickolls, I.~Buck, M.~Garland, and K.~Skadron, ``Scalable parallel
  programming with {CUDA},'' {\em Queue}, vol.~6, no.~2, pp.~40--53, 2008.

\bibitem{kingma2014adam}
D.~Kingma and J.~Ba, ``Adam: A method for stochastic optimization,'' {\em arXiv
  preprint arXiv:1412.6980}, 2014.

\bibitem{dropout}
N.~Srivastava, G.~Hinton, A.~Krizhevsky, I.~Sutskever, and R.~Salakhutdinov,
  ``Dropout: A simple way to prevent neural networks from overfitting,'' {\em
  Journal Machine Learning Research}, vol.~15, no.~1, pp.~1929--1958, 2014.

\bibitem{synapticnoise}
K.-C. Jim, C.~L. Giles, and B.~G. Horne, ``An analysis of noise in recurrent
  neural networks: convergence and generalization,'' {\em IEEE Trans. Neural
  Netw.}, vol.~7, no.~6, pp.~1424--1438, 1996.

\bibitem{fang2013classifying_performance_measures}
Y.~Fang, X.~Wang, E.~K. Michaelis, and J.~Fang, ``Classifying aging genes into
  {DNA} repair or non-{DNA} repair-related categories,'' {\em International
  Conference Intelligent Computing}, pp.~20--29, 2013.

\end{thebibliography}
	\end{spacing}
	}
\end{document}